# Towards a Privacy and Security-Aware Framework for Ethical AI: Guiding the Development and Assessment of AI Systems

Daria Korobenko, University of Tartu, Estonia
Anastasija Nikiforova, University of Tartu, Estonia
Rajesh Sharma, University of Tartu, Estonia

**Abstract**: As artificial intelligence (AI) continues its unprecedented global expansion, accompanied by a proliferation of benefits, an increasing apprehension about the privacy and security implications of AI-enabled systems emerges. The pivotal question of effectively controlling AI development at both jurisdictional and organizational levels has become a prominent theme in contemporary discourse. While the European Parliament and Council have taken a decisive step by reaching a political agreement on the EU AI Act, the world's first comprehensive AI law, organizations still find it challenging to adapt to the fast-evolving AI landscape, lacking a universal tool for evaluating the privacy and security dimensions of their AI models and systems. In response to this critical challenge, this study conducts a systematic literature review (SLR) spanning the years 2020 to 2023, with a primary focus on establishing a unified definition of key concepts in AI Ethics, particularly emphasizing the domains of privacy and security. Through the synthesis of knowledge extracted from the SLR, this study presents a conceptual framework tailored for privacy- and security-aware AI systems. This framework is designed to assist diverse stakeholders, including organizations, academic institutions, and governmental bodies, in both the development and critical assessment of AI systems. Essentially, the proposed framework serves as a guide for ethical decision-making, fostering an environment wherein AI is developed and utilized with a strong commitment to ethical principles. In addition, the study unravels the key issues and challenges surrounding the privacy and security dimensions, delineating promising avenues for future research, thereby contributing to the ongoing dialogue on the globalization and democratization of AI ethics.

## 1 INTRODUCTION

Artificial intelligence (AI) is becoming increasingly prevalent in contemporary society, permeating various aspects of our daily lives, from virtual assistants such as Siri and Alexa to the deployment of AI in autonomous vehicles and facial recognition systems. This increasing integration of AI technologies into everyday routines reflects a broader societal acceptance and underscores its movement into the mainstream and prompting major industry players, including Microsoft, Google, and Meta, to release their first publicly accessible generative AI applications contributing to a transformative paradigm within the field of Artificial Intelligence [37].

As Artificial Intelligence continues to advance, it has brought forth a myriad concerns regarding the ethical aspects related to the collection, processing, and storage of personal data [52]. AI technologies require extensive datasets to refine their algorithms and enhance performance, often involving personal and sensitive information such as medical records and social security numbers. A primary concerns surrounding AI pertains to the risk of unauthorized access to personal data and potential security breaches. The vast amounts of data collected and processed create an elevated threat of vulnerability to security attacks. In particular, cybercrimes impact the security of 80% of global businesses, highlighting the profound consequences that mishandled personal data can unleash [3]. Additionally, concerns arise from the application of AI in surveillance, sparking debates over privacy rights and the potential for these technologies to be misused [21].

The problem becomes even more evident with an increasing number of adverse incidents within the realm of AI. Notable examples include Facebook's data privacy scandal in 2018 when it was revealed that Cambridge Analytica, a political consulting firm, harvested personal data from millions of Facebook users without their consent [14]. Similarly, Google's mishandling of medical data in 2019 through Project Nightingale, where sensitive patient information was accessed without proper authorization [43], further spotlighted the criticality of privacy and security in the AI landscape. Nevertheless, a more extensive array of concerns may exist, some of which are possibly undisclosed to the public.

This underscores the need for a balanced and ethical approach to the development and deployment of AI, emphasizing the importance of safeguarding personal data and respecting privacy rights in the face of advancing technological capabilities. Researchers worldwide have devoted considerable efforts to exploring these inquiries, with discussions originating in the early 2000s [38]. But despite the ongoing dialogue, no consensus has been reached regarding key terms and concepts around AI Ethics [46], and there is currently no well-defined framework for evaluating the privacy and security of AI models. In a significant development, the European Union Parliament, after months of extensive negotiations, reached an agreement on the first legal framework dedicated to AI - the EU AI Act [4]. The introduction of this framework has sparked widespread (scientific) discourse, critically examining its quality and applicability [25].

In light of these developments and the recognized gap in the field, this study aims to perform an analysis of recent AI Ethics literature using a Systematic Literature Review (SLR) to uncover key themes and knowledge gaps in the privacy and security domains. Subsequently, our objective is to conceptualize the results in a security and privacy-aware framework for designing and evaluating AI models.

The following research questions have been formulated to attain the central objectives of this study:

- ***RQ1:*** *What is the current state of the art of AI ethics academic literature focused on privacy and security perspectives?*
- ***RQ2:*** *What are the existing frameworks for assessing the adherence of an AI model to the AI ethics principles of privacy and security?*
- ***RQ3:*** *What aspects should be considered to ensure the development and deployment of privacy- and security-aware AI model?*

The remainder of the paper is structured as follows: Section 2 presents the background of the study, Section 3 presents the research methodology. Section 4 discusses the SLR results, while Section 5 presents a framework. Finally, Section 6 provides an overview of threats to the validity of the study and limitations, and Section 7 presents findings and defines future directions.

## 2 THEORETICAL BACKGROUND

Artificial Intelligence (AI) ethics has evolved into a pivotal field of study, gaining prominence as AI technologies become increasingly intertwined with our daily lives. The intricate nature of defining AI ethics is reflected in the myriad of contradicting and overlapping interpretations, underscoring the complexity of ethical considerations in AI development and deployment.

While some studies delve into the ethical considerations of AI creators, such as morality of AI, highlighting the ethical obligations of those involved in designing and developing AI systems [7], others define AI Ethics others define AI Ethics within a broader context, examining intricate connections with domains such as information ethics, computer ethics, and robot ethics [22]. These studies shed light on the multifaceted nature of ethical concerns in the digital and automated world, showcasing the overlapping and interdependent relationships between these fields. However, a predominant focus in this domain revolves around delineating AI Ethics based on a set of fundamental principles [10]. Ethical principles offer a structured approach to comprehending and navigating the ethical intricacies inherent in AI technology and its applications. Notably, the work by Jobin et al. is considered pivotal in this context [5, 50], providing the most comprehensive analysis of principles in ethical AI guidelines to date. Through the analysis of 84 ethical AI guidelines, 11 central principles were identified, including transparency, justice and fairness, non-maleficence, responsibility, privacy, beneficence, freedom and autonomy, trust, dignity, sustainability, and solidarity [29].

Privacy and security principles, often conceptually linked together in scientific literature [29], stand out among the AI ethics principles due to their fundamental role in safeguarding individual rights and societal well-being [30]. Privacy ensures the protection of personal data from unauthorized access and use, while security addresses the integrity and robustness of AI systems, protecting against unauthorized access and mitigating risks of malicious interference. The major public concerns in this regard are centered around the privacy and security vulnerabilities in AI systems controlling autonomous vehicles and healthcare data breaches [13], showcasing the broader implications of neglecting these principles. Hacks and manipulations in these systems not only pose risks to individual safety but also threaten public well-being. This further emphasizes that addressing privacy and security is not just an ethical imperative but essential for fostering public trust and ensuring the responsible development and deployment of AI technologies.

While the significance of privacy and security in the realm of AI Ethics is undeniable, the existing scientific literature lacks a practical framework to address these issues comprehensively. Several general AI ethics frameworks have been proposed, reflecting a collective desire to establish a universal standard that ensures the responsible and ethical use of AI. Frameworks like the Ethics Guidelines for Trustworthy AI [1] have commendably strived to provide guidelines for ethical AI development. The definition provided by the European High-Level Expert Group on "Trustworthy AI" (TAI) encompasses three fundamental dimensions: (1) being lawful, i.e., ensuring compliance with all applicable laws and regulations; (2) being ethical, by upholding adherence to ethical principles and values; and (3) being robust, both technically and socially. However, a noticeable gap exists in the depth to which ethical principles are embedded in these frameworks. They often offer high-level guidance, leaving room for interpretation and potential oversights in implementing concrete measures to safeguard privacy and enhance security in AI systems.

Regulatory bodies, notably the European Union, have recognized the imperative to address the ethical and societal impact of AI, thus proposing EU AI Act - the first regulation on artificial intelligence [2]. The framework follows a risk-based approach to classify AI systems into 4 potential risk levels: (1) unacceptable risk, (2) high risk, (3) limited risk, (4) minimal or no risk. The framework aims to define clear requirements and obligations specific to each risk level for AI systems and its creators. While commendable strides have been made in formulating it, existing framework has a number of limitations and gaps [25], further emphasizing the critical need for a holistic approach that integrates privacy and security considerations.

The urgency of this topic is underscored by the rapid advancement of AI technologies and their pervasive integration into diverse sectors. The ethical implications and societal consequences of AI require a proactive approach, calling for frameworks that not only uphold ethical principles but also embed security and privacy considerations. Therefore, this study seeks to propose a privacy- and security-aware framework for Ethical AI, by offering a more granular and explicit integration of privacy and security principles into the broader ethical landscape of AI.

## 3 RESEARCH METHODOLOGY

To achieve a more profound understanding of privacy and security domains in AI Ethics, forming the foundation for the framework development, this study adopts a systematic literature review (SLR) approach, as defined by Kitchenham [35]. Unlike narrative literature reviews, the SLR approach is an evidence-based approach characterized by a transparent, rigorous, and replicable process for literature searching and review. The review process encompasses the following key steps: 1) study identification, 2) study selection, 3) assessment of study relevance and quality, 4) data extraction, and 5) data synthesis. This systematic approach ensures a comprehensive and well-structured analysis of existing literature, contributing to a robust foundation for the development of the privacy- and security-aware framework for Ethical AI.

### 3.1 Study Identification

The primary aim of this initial phase is to comprehensively locate relevant studies related to the research questions through an unbiased search strategy. To achieve this goal, Scopus, alongside Web of Science, stands as one of the most esteemed scientific databases, offering extensive collections of academic resources. Notably, Scopus is recognized as the largest global indexer of research output, with over 7000 publishers in 105 countries [41]. As the most extensive interdisciplinary abstract and citation database, Scopus provides a more comprehensive resource for conducting an in-depth literature review. Therefore, it has been used for this SLR. The set of search keywords (see Table 1) was defined to attain the objective of unveiling key themes and knowledge gaps within the privacy and security domains of AI Ethics. The search process underwent multiple iterations, with the final search conducted on January 1, 2024.

**Table 1.** Search terms used for the literature review

| Database | Search terms in title/abstract/keywords |
|---|---|
| Scopus | ("AI ethics" OR "artificial intelligence ethics") AND ( "privacy" OR "security") |

### 3.2 Study Selection

In the second step, the selection of studies, we defined the exclusion and inclusion criteria (Table 2).

Since AI is evolving at a fast pace, we limited our selection to peer-reviewed studies published in the last three years, i.e., between January 1, 2020, and December 31, 2023, and written in the English language. The defined exclusion criteria filter the search string findings and remove irrelevant, not accessible, redundant, and low-quality studies.

**Table 2.** Inclusion and exclusion criteria

| **Number** | **Inclusion criteria** |
|---|---|
| In1 | Studies published in 2020-2023 |
| In2 | Studies written in English language In3 |
| | Peer-reviewed studies |
| **Number** | **Exclusion criteria** |
| Ex1 | Books, book chapters, and conference reviews Ex2 |
| | Studies that are not available in full text |
| Ex3 | Studies that are not focused on AI Ethics, specifically privacy or security |

## 3.3 Study Relevance and Quality Assessment

The third step of our SLR involved an assessment of the relevance and quality of the selected studies. First, each identified study underwent scrutiny based on its title and abstract to filter out studies that did not explicitly focus on AI Ethics, particularly on the privacy or security. Subsequently, the remaining studies underwent an evaluation of their relevance and quality through an examination of the full text.

The relevance assessment was conducted based on a predefined set of specific criteria that revolved around the central role and focus of AI Ethics. A primary consideration was whether AI Ethics constituted a substantial or major component of the study, evident through its research questions, objectives, and overarching thematic focus. Furthermore, the inclusion of privacy or security topics within the study was essential for it to be deemed relevant. Studies that did not, at the very least, partly focus on these aspects or treated them as peripheral issues were excluded during this phase of the assessment.

Building upon this screening, we further scrutinized each study using the following quality dimensions [53]:

- the objectives of the study are clearly stated, and data collection methods are adequately described. References support important statements in the paper;
- the design of the study is appropriate for the research objectives. The study's research questions are answered or the research objective is attained;
- the study's research approach is described in sufficient detail.

Figure 1 visualize the entire process of study screening and selection, using PRISMA (Preferred Reporting Items for Systematic Reviews and Meta-Analyses) flowchart. The final selection led us to only 73 studies directly addressing the privacy and security aspects of AI.

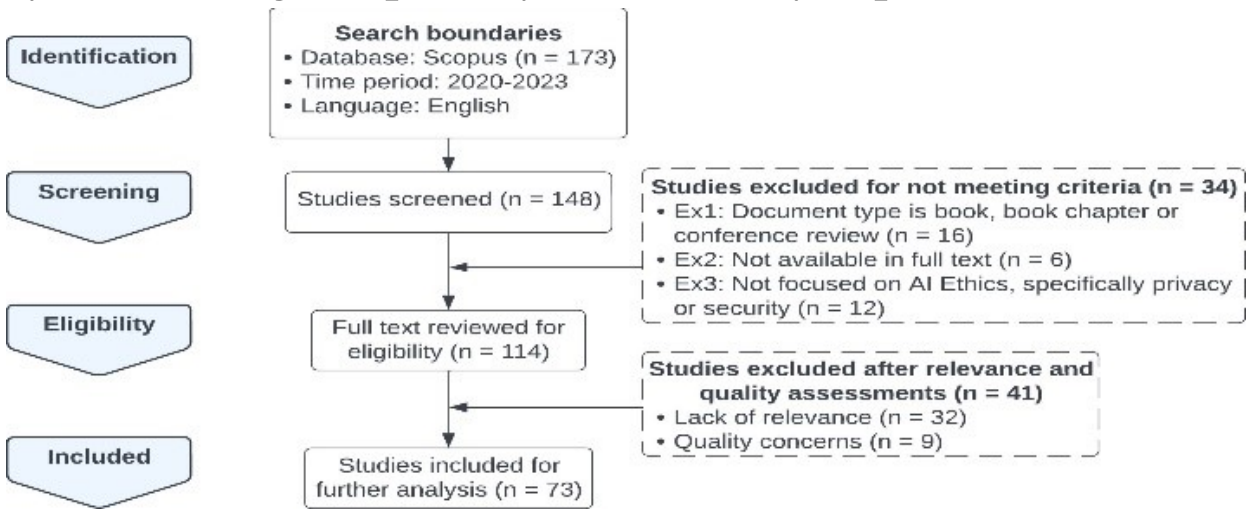

**Fig. 1.** Study selection, assessment, and inclusion (presented using the PRISMA flow diagram)

### 3.4 Data Extraction

For the data extraction phase of our SLR, a protocol was created to record the metadata for each of the selected studies. Table 3 shows the structure of the protocol with the metadata we collected on all 73 studies, including (1) descriptive information, (2) approach-related information, and (3) AI-related information. To ensure alignment with our objectives, these metadata categories were derived from the research questions:

- **descriptive information** captures essential details about each study, such as authors, publication year, and title that provides a foundational overview to identify the context of the research;
- **approach-related information** focuses on the methodologies employed in the selected studies, exploring the research approaches and data collection methods, which helps to assess the reliability and validity of the findings;
- **AI-related information** delves into AI-specific details given the focus on AI in the selected study. It includes information about the types of AI techniques utilized and noteworthy findings related to AI Ethics.

**Table 3.** Overview of information collected about each of the selected studies

| Category | Metadata | Description |
|---|---|---|
| **Descriptive information** | # | What is the study number, assigned in an Excel worksheet? |
| | Title | What is the study title? |
| | Author(s) | Who is/are the study's author(s)? |
| | Year | In which year was this study published? |
| | Country | Which country is/are the author(s) affiliated with? |
| | Document Type | What is the type of this study (e.g. journal article)? |
| | Publication | Where was this study published (e.g. name of the journal/conference)? |
| | DOI / Website | What is the study's DOI? If no DOI is available, through which website can this study be found? |
| | Keywords Citation | What are the keywords in this study? What is the citation count for this study? |
| **Approach-related** | Objective | What is the research objective / main question? |
| | Research method | Which method was used to collect data in the study? |
| | Approach | Does the study use qualitative, quantitative or mixed approach? |
| | Availability of underlying research data | Does the study contain a reference to the public availability of the underlying research data (or explains why this data is not openly shared)? |
| | Relevance | Does the study's research focus align with the objectives of this review? |
| | Quality concerns | Are there any quality concerns (e.g., limited information about the research methods used)? |
| | Primary domain | What is the primary domain/field in this study (e.g., healthcare)? |
| | Secondary domain | What is the secondary domain/field in this study (e.g., education)? |
| **AI-related information** | Technology | Which technology is this article primarily focused on (e.g., machine learning)? |
| | AI ethics | How AI ethics is defined in the article? |
| | Ethical principles | What ethical principles are mentioned in the article (e.g., transparency)? |
| | Emphasis on privacy or security | Does the study place a stronger emphasis on privacy or security, or both are considered equally? |
| | Framework | Does the study incorporate any specific framework, and if so, which one? |
| | Framework application phase | What is the application phase of the framework (if applicable)? |
| | Key findings | What are the key findings in this study? |

### 3.5 Data Synthesis

The data synthesis phase, constituting the final step of our SLR, involved a thorough analysis of the previously extracted metadata. To this end, we utilized the Python programming language (mainly the Matplotlib library) to summarize and visualize our findings, elucidating key data patterns and trends (Section 4). Building upon these insights, we subsequently developed a privacy- and security-aware framework for Ethical AI, as detailed in Section 5. This framework stands a direct outcome of the conducted SLR, providing a nuanced approach to understand and address privacy and security concerns in AI Ethics.

## 4 RESULTS OF THE SYSTEMATIC LITERATURE REVIEW

Let us present the results obtained from the analysis of 73 selected research articles that concern privacy and security domains of AI Ethics. The data underlying this study is publicly available in Zenodo, DOI: https://doi.org/10.5281/zenodo.10451282; thereby supporting the open science movement and ensuring replicability of the study.

### 4.1 Descriptive Analysis

Descriptive Analysis provides the general overview on the examined studies and includes the examination of chronological publication trends, analysis of the publication types, overview of the geographical landscape, citation and keyword analysis.

*4.1.1 Chronological Publication Trends.* The years 2020 to 2023 witnessed a noteworthy upward in AI Ethics research, evident in the escalating number of publications indexed in Scopus. Commencing with a modest count of 4 publications in 2020, the figures experienced a substantial ascent, reaching a peak of 32 by 2023, as shown in Figure 2. This trend underscores the growing prominence of AI Ethics as a research domain and underscores the amplified scholarly focus on the ethical dimensions associated with artificial intelligence during this period.

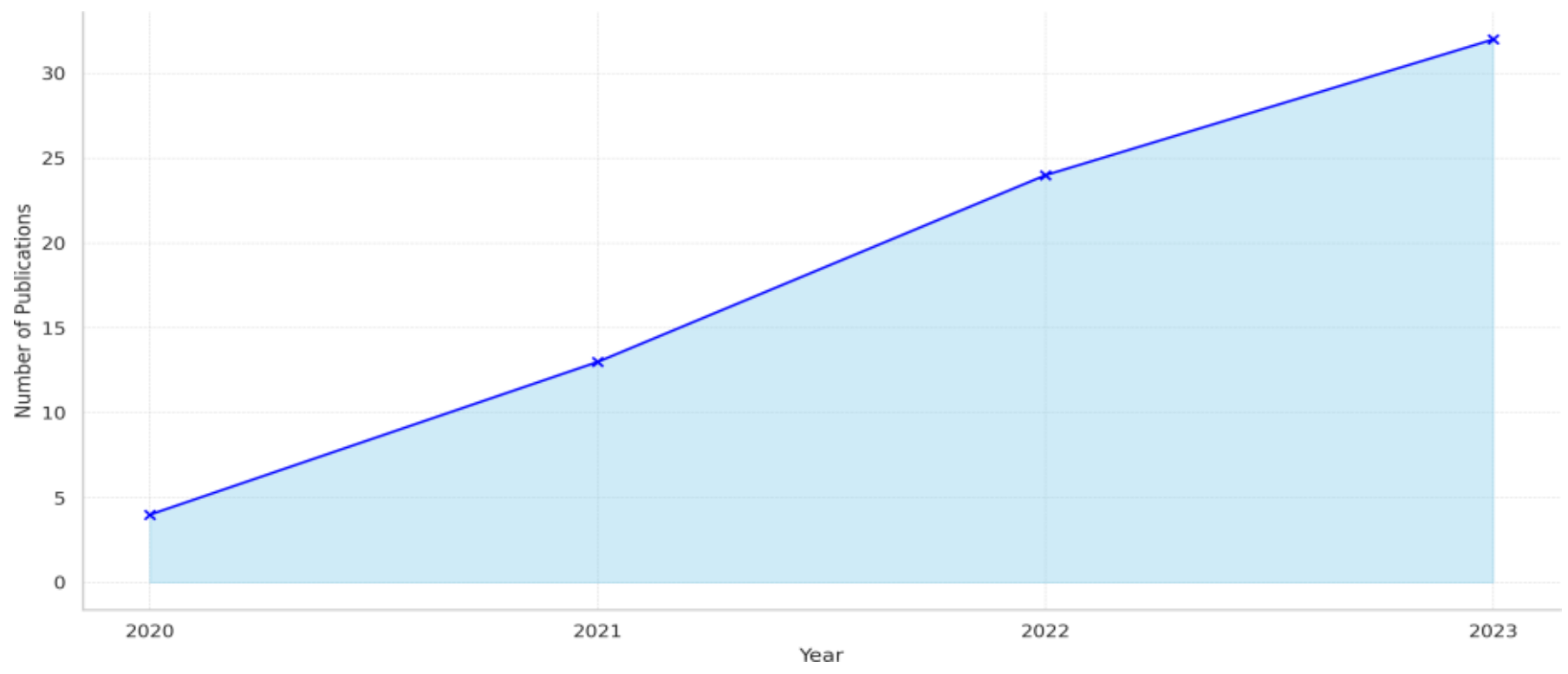

Fig. 2. Publication trends over time

*4.1.2 Publication Type Distribution.* The resulting dataset underscores a focused trajectory in scholarly output, predominantly featuring journal articles and conference papers, as shown in Figure 3. Journal articles, accounting for approximately 67.12% of the studies, are the prevalent publication type, showing the field's strong emphasis on rigorous peer-reviewed research. Conference papers account for roughly 32.88%, reflecting the field's vibrant exchange of ideas and findings in dynamic, collaborative events, such as conferences and forums.

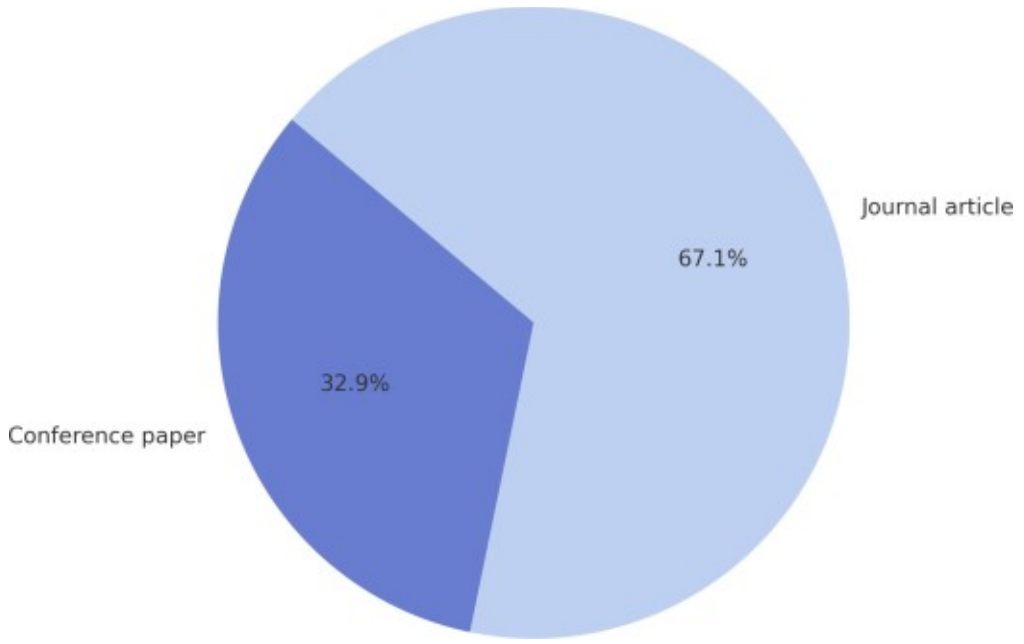

**Fig. 3.** Publication type distribution

It should be noted that while our initial dataset included a broader spectrum of publication types, such as books, book chapters, and conference reviews, these were methodically filtered out during the screening process. The exclusion of these publication types does not detract from their value but rather serves to sharpen the focus of our current inquiry.

*4.1.3 Geographical Landscape.* The data provides a clear picture of the global research landscape in the field of AI Ethics, emphasizing countries (Figure 4) that are key contributors in the scientific arena.

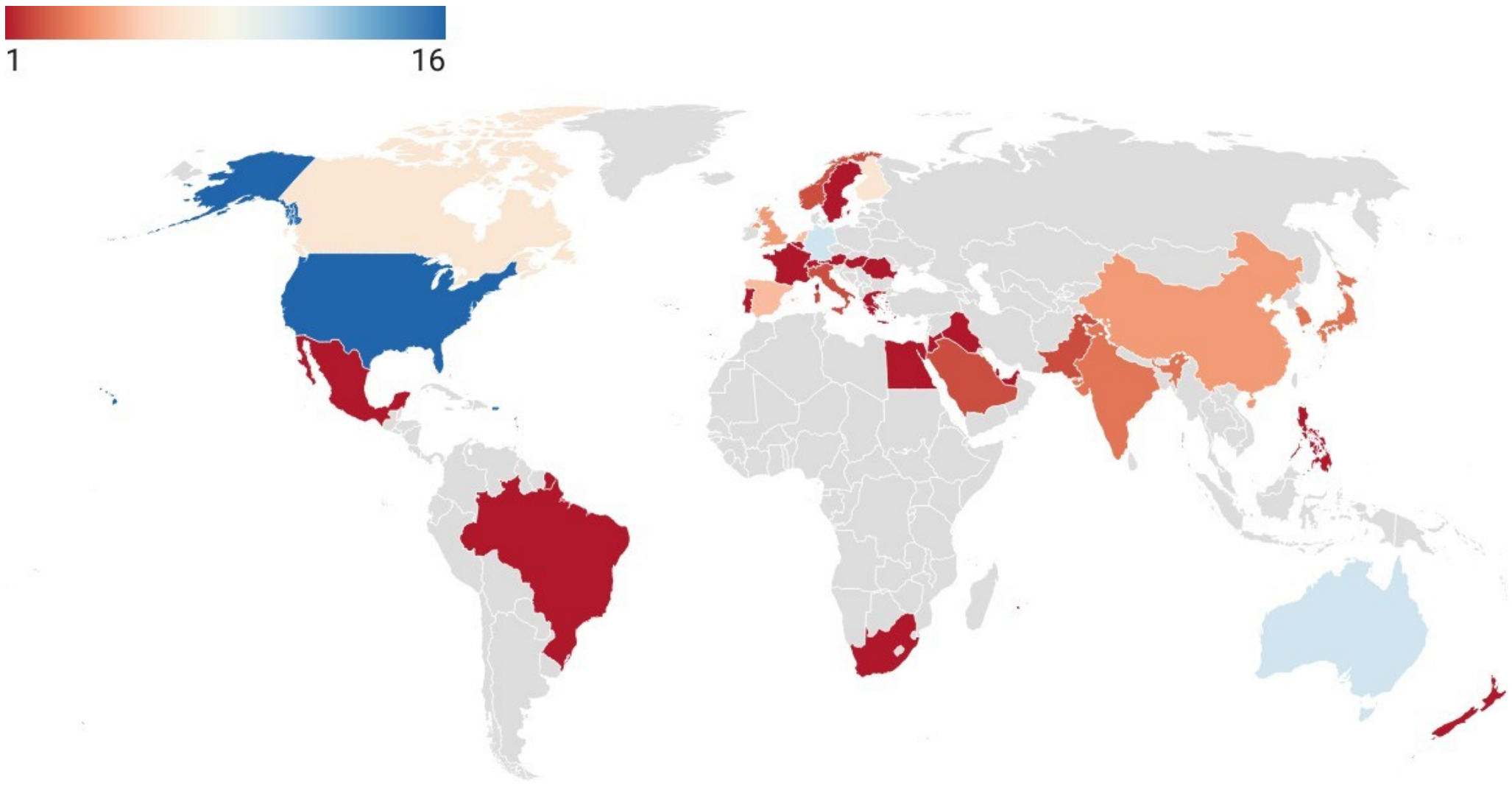

**Fig. 4.** Geographical distribution of publications

The United States secures a leading position in our dataset, contributing 15.09% of the total research publications. This substantial share underscores the strong academic and research infrastructure in the United States. Australia and Germany, each with a notable contribution of 10.38%, reflect their considerable involvement in AI Ethics research. This is especially noteworthy for Australia, which maintains a high output despite its smaller population size compared to the United States and Germany. Canada and Finland, contributing 6.6% each, also highlight the importance and focus on AI Ethics research within these countries, indicating a strong commitment to advancing in this domain. Other countries like the Netherlands, Spain, China, and the United Kingdom, each offering around 4-5% of the total publications, further demonstrate the global interest in AI Ethics.

From a regional perspective, Europe stands out as the dominant force, accounting for 41.51% of the total research output. This significant percentage suggests a dense concentration of research activities spread across several European countries, marking the continent as a central hub in the field of AI Ethics. North America, led by the United States and supported by Canada, also plays a crucial role, with a combined contribution of 22.64%, reflecting the region's strong influence in shaping global research trends. Asia, represented by countries like China, India, Japan, and the Republic of Korea, contributes 19.81%, showcasing the region's growing momentum in AI Ethics research, driven by technological advancements and a focus on innovation. Oceania, predominantly represented by Australia, makes a significant contribution as well, with 11.32% of the total publications, indicating its active role in the global AI Ethics research landscape. While less frequent, contributions from Africa and South America are essential for adding diversity to the global research perspectives.

In summary, these findings highlight a widespread engagement in AI Ethics research, with notable concentrations in Europe and North America, and growing participation from regions like Asia and Oceania.

*4.1.4 Citation Analysis.* Citation analysis presented in Figure 5 shows a skewed distribution with a majority of articles receiving a lower number of citations, while a few have significantly higher citations. This pattern is typical in academic literature, where certain pioneering or breakthrough studies gain substantial recognition. This can be also attributed to many studies being relatively recent and not having sufficient time to accumulate citations. Additionally, the zero-citation count for 30.13% of articles may reflect the time lag in the dissemination and integration of new research into the broader academic discourse.

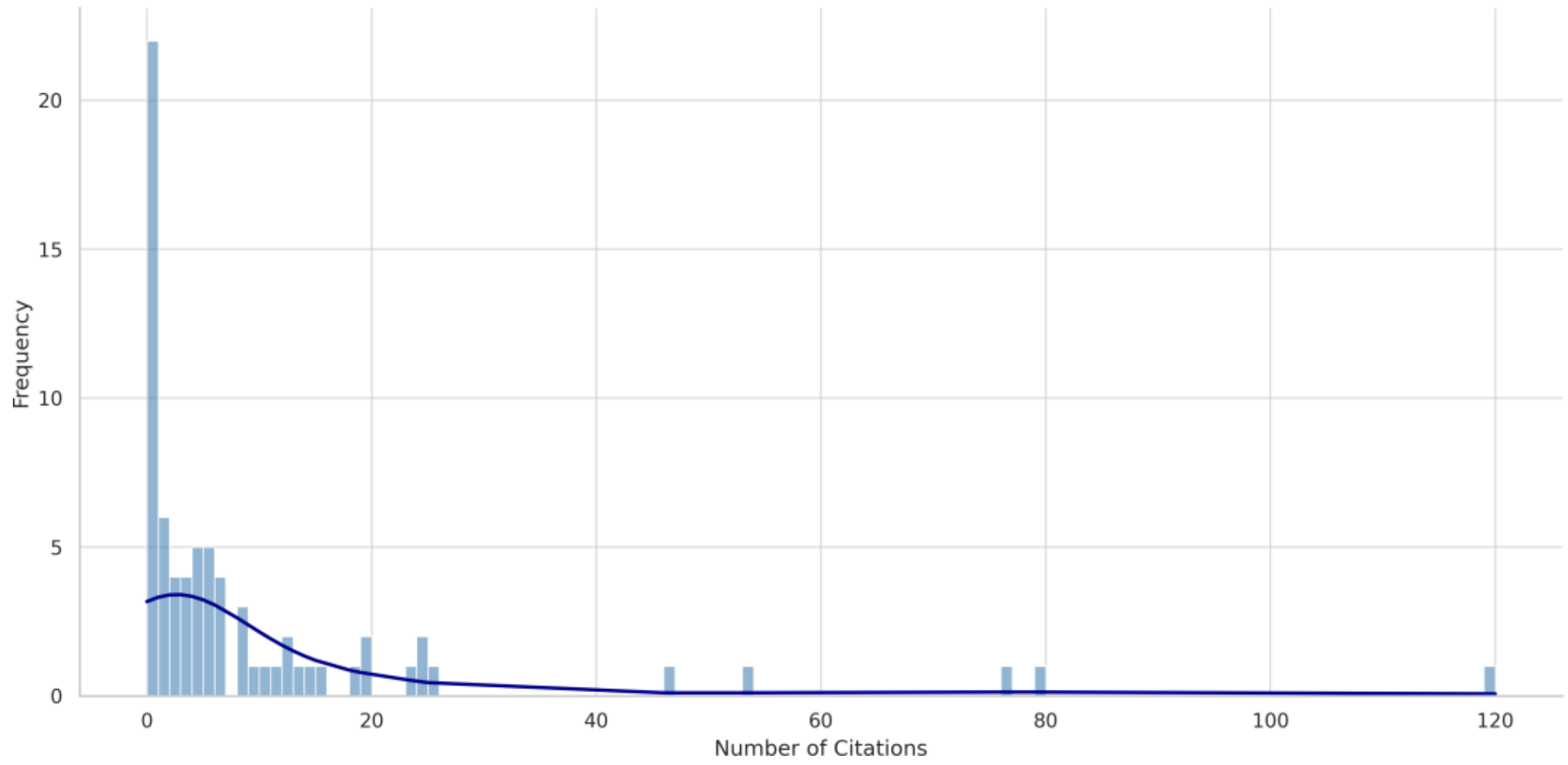

**Fig. 5.** Citation analysis

*4.1.5 Keyword Analysis.* A word cloud visualization of keywords is employed to present the predominant themes emerging from the dataset, as shown in Figure 6.

A word cloud offers an intuitive and visually engaging means to understand the frequency of keywords within a text corpus. In this graphical representation, the size of each word or phrase is directly proportional to its frequency of occurrence. Therefore, more prominent words in the word cloud, excluding those utilized in our search queries, are "*Machine learning*" and "*Machine ethics*", "*Trustworthy AI*", "*AI canvas*" and "*Responsible AI*", "*Data governance*" and "*AI governance*", "*AI Act*" and "*GDPR*", and various AI Ethics principles. The dominance of these particular words underscores their significance in the context of our research.

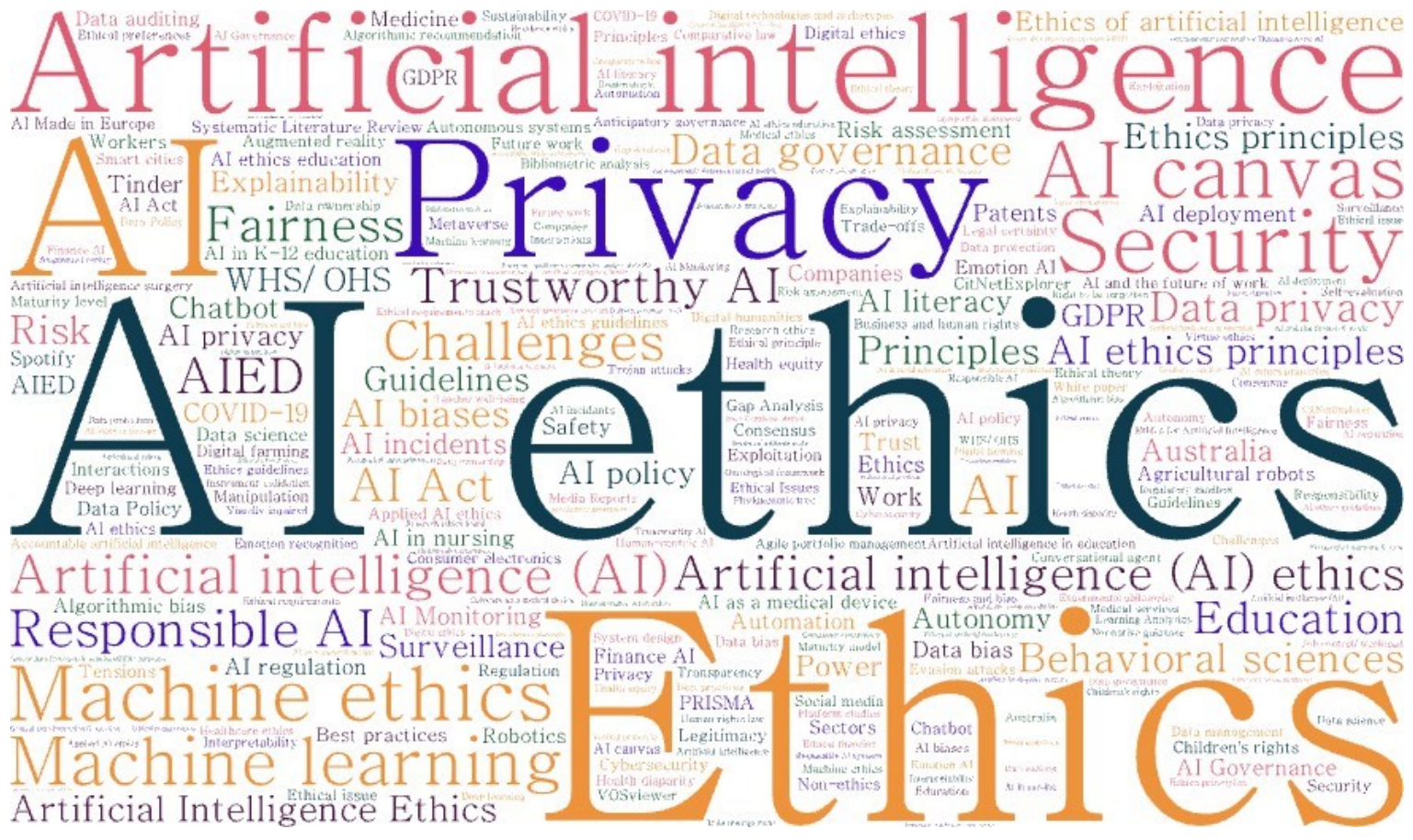

**Fig. 6.** Word cloud based on keywords

## 4.2 Approach Analysis

The selected studies utilize a variety of research methods, with literature review being by far the most dominant research approach (n = 40) (Figure 7). The array of other methods employed in the examined studies includes document analysis, surveys, interviews, case studies, expert panels, comparative analysis, workshops, action research, and review of existing initiatives.

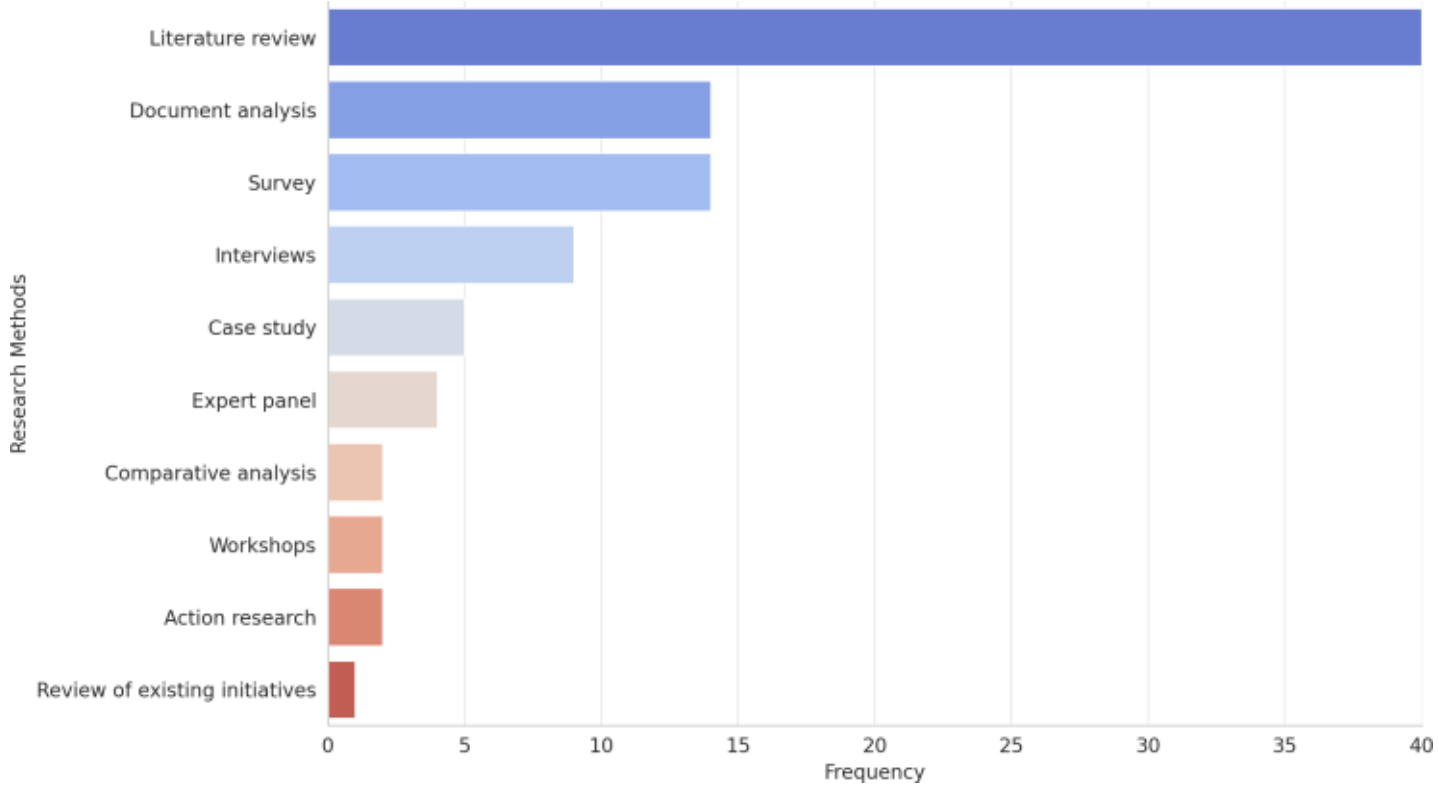

Fig. 7. Research methods utilized in publications

83.56% of the studies in our sample utilize qualitative method, and the remainder of the studies employ quantitative or mixed methods (combining qualitative and quantitative approaches) (Figure 8).

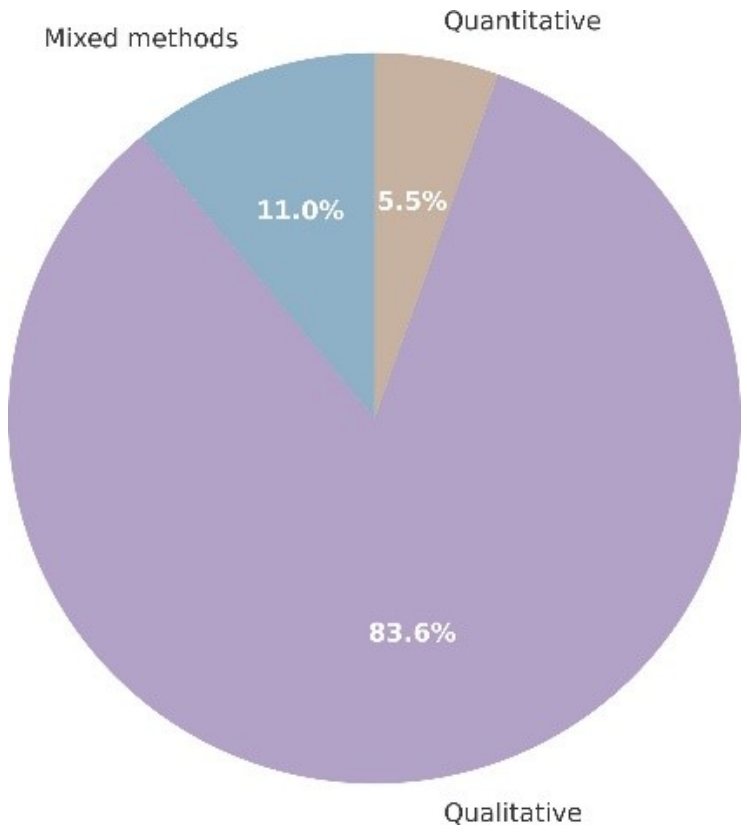

**Fig. 8.** Research approaches utilized by studies

Only 12 studies have openly made the underlying research data available, despite the growing trend in openly sharing the underlying research data as a positive open science practice, which increases transparency and trust and allows scrutiny of the findings [53]. 3 studies mention that research data can be made available upon request from the author.

## 4.3 Quality Analysis

For 43 out of the 73 studies, the research design was appropriate. For 30 studies, we had minor quality concerns – for instance, missing information about the literature review methodology, such as the total number of search results for each database explored or the quality assessment mechanisms of the reviewed studies. Studies for which we had significant quality concerns (n = 9) had already been removed during the full study assessment.

### 4.4 Content Analysis

*4.4.1 AI Ethics Definition and Principles.* Out of the 73 studies that were reviewed, only 7 included a formal definition of AI Ethics, as shown in Table 4. While several of these studies acknowledged the absence of a scientific consensus on the definition of AI Ethics [46], the predominant approach was to define AI Ethics in terms of its principles. For instance, Ashok et al. [10] described AI Ethics by highlighting its core principles of responsibility, transparency, auditability, incorruptibility, predictability, and the morality of machines.

Table 4. AI Ethics Definitions

| **AI Ethics Definition** | **Reference** |
|---|---|
| *"deals with the issues raised in developing, deploying, and using AI systems and involves the moral behavior of humans in the design and development of AI"* | [6] |
| *"a human-centered perspective which focuses on the morality of humans who deal with the AI systems, including developers, manufacturers, operators, consumers, etc."* | [7] |
| *"the moral behavior of humans as well as the moral behavior of AI agents in the process of designing, constructing, using, and handling AI beings"* | [12] |
| *"a set of values, principles, and techniques that employ widely accepted standards of right and wrong to guide moral conduct in the development and use of AI technologies"* | [17] |
| *"an emerging and interdisciplinary field concerned with addressing ethical issues of AI. AI ethics involves the ethics of AI, which studies the ethical theories, guidelines, policies, principles, rules, and regulations related to AI, and the ethical AI, that is, the AI that can uphold ethical norms and behaves ethically"* | [26] |
| *"a subfield of applied ethics focusing on the ethical issues raised in the development, deployment, and use of AI"* | [27] |
| *"specifies the moral obligations and duties of an AI and its creators"* | [49] |

Subsequently, we identified the key principles of AI Ethics prevalent in the studies (Figure 9). The discourse is heavily dominated by principles of **transparency**, **privacy**, and **fairness**, whereas the principle of sustainability, among others, garners considerably less scholarly attention. The principles of beneficence and non-maleficence, while being significantly covered in the literature, might be less emphasized due to their broader definitions and the challenges associated with their practical implementation.

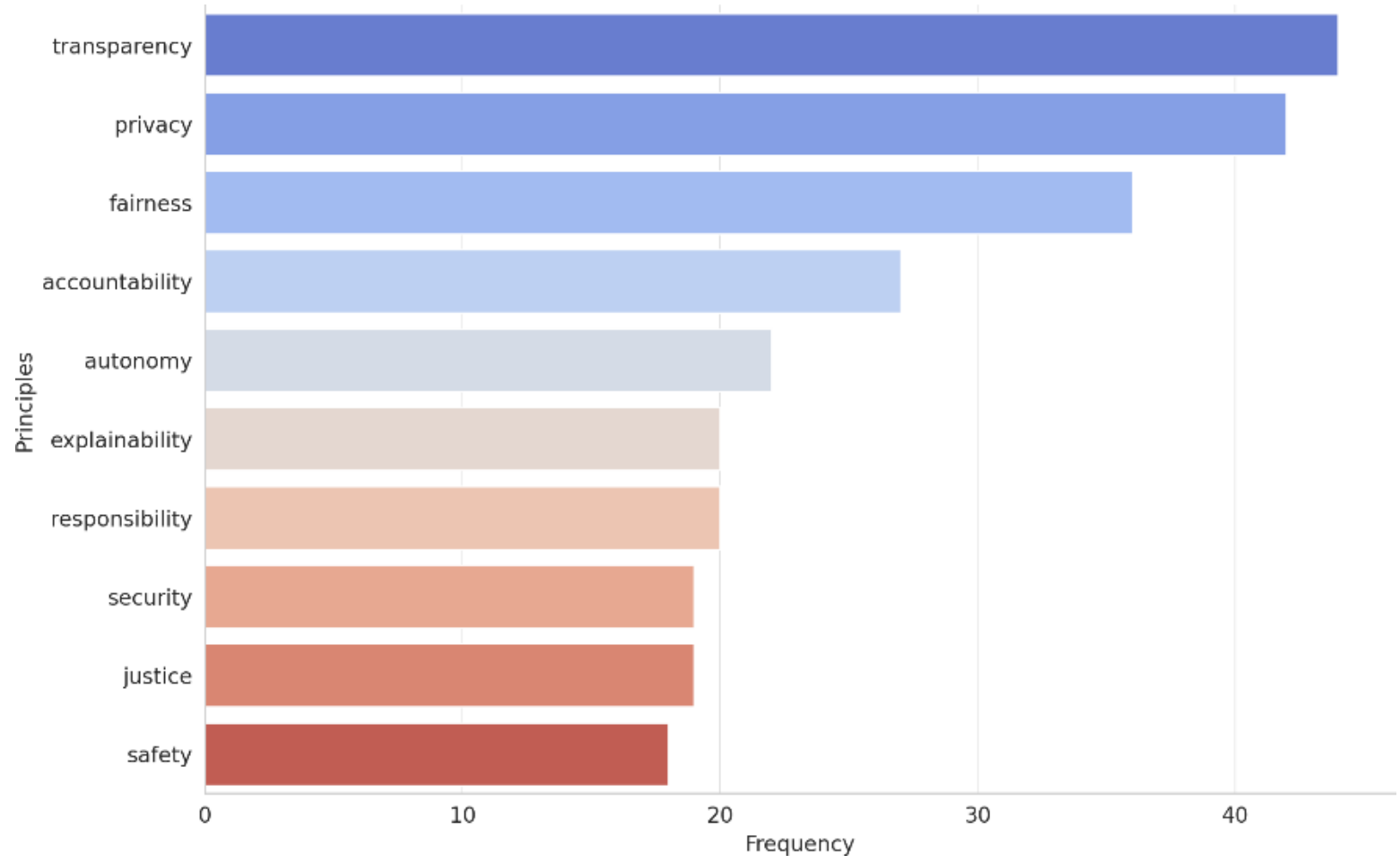

Fig. 9. Ethical principles identified in literature

By synthesizing the principles that exhibit interrelation and conceptual linkage in the literature, we have identified the five most frequently cited ethical principles, namely **transparency** and **explainability**, **privacy** and **security**, **fairness** and **justice**, **responsibility** and **accountability**, and **freedom** and **autonomy**. A brief description of each of these principles is provided in Table 5. The ordering of these principles in the table follows a top-down order, reflecting their popularity in the current scientific discourse.

**Table 5.** AI Ethics Principles

| Ethical principle | Description |
|---|---|
| Transparency and explainability | AI systems should be designed to provide clear insight into their processes and decisions, ensuring that users and regulators can understand AI actions and outputs |
| Privacy and security | AI systems should safeguard personal data, respecting privacy rights and maintaining robust defenses against unauthorized access and breaches |
| Fairness and justice | AI systems should be inclusive and operate without bias, offering equal opportunity and treatment while ensuring just outcomes for all individuals and groups |
| Responsibility and accountability | There should be a clear assignment of responsibility for AI behavior, with mechanisms in place to hold the creators accountable for the system's impact |
| Freedom and autonomy | AI systems should enhance human decision-making without constraining individual freedoms, allowing for personal autonomy and self-determination |

As a result of the SLR, we define AI Ethics as ***an interdisciplinary field focused on the moral conduct of both humans and AI agents involved in the development, deployment, and use of AI technologies***. Central to AI Ethics are the principles of **transparency** and **explainability**, **privacy** and **security**, **fairness** and **justice**, as well as **freedom** and **autonomy**.

*4.4.2 Classification of Research Articles.* In our investigation of the interdisciplinary landscape of AI Ethics, we performed a domain analysis for the selected studies to map out the diverse fields entwined with AI Ethics.

Each study underwent thorough scrutiny to identify its primary domain, which signifies the principal area of research or application. The primary domain is the central focus of a study, wherein

AI ethics is predominantly addressed or scrutinized. For instance, if a study primarily addresses the data privacy of patients, *Healthcare* would be assigned as its primary domain.

Furthermore, we also sought to identify secondary domains represented by each study, which is an auxiliary field that provides additional context, background, or implications for the primary discussion of AI ethics. It supports the primary domain but is not the main focus of the ethical inquiry. For example, a study primarily focused on *Healthcare* might also touch upon the current governance initiatives aimed at addressing privacy issues, thus assigning *Law and Governance* as a secondary domain.

The domain analysis is visualized in a network graph (Figure 10), where nodes represent domains and edges signify the conceptual links between these domains and AI Ethics. The size of the node corresponds to the number of studies within the domain.

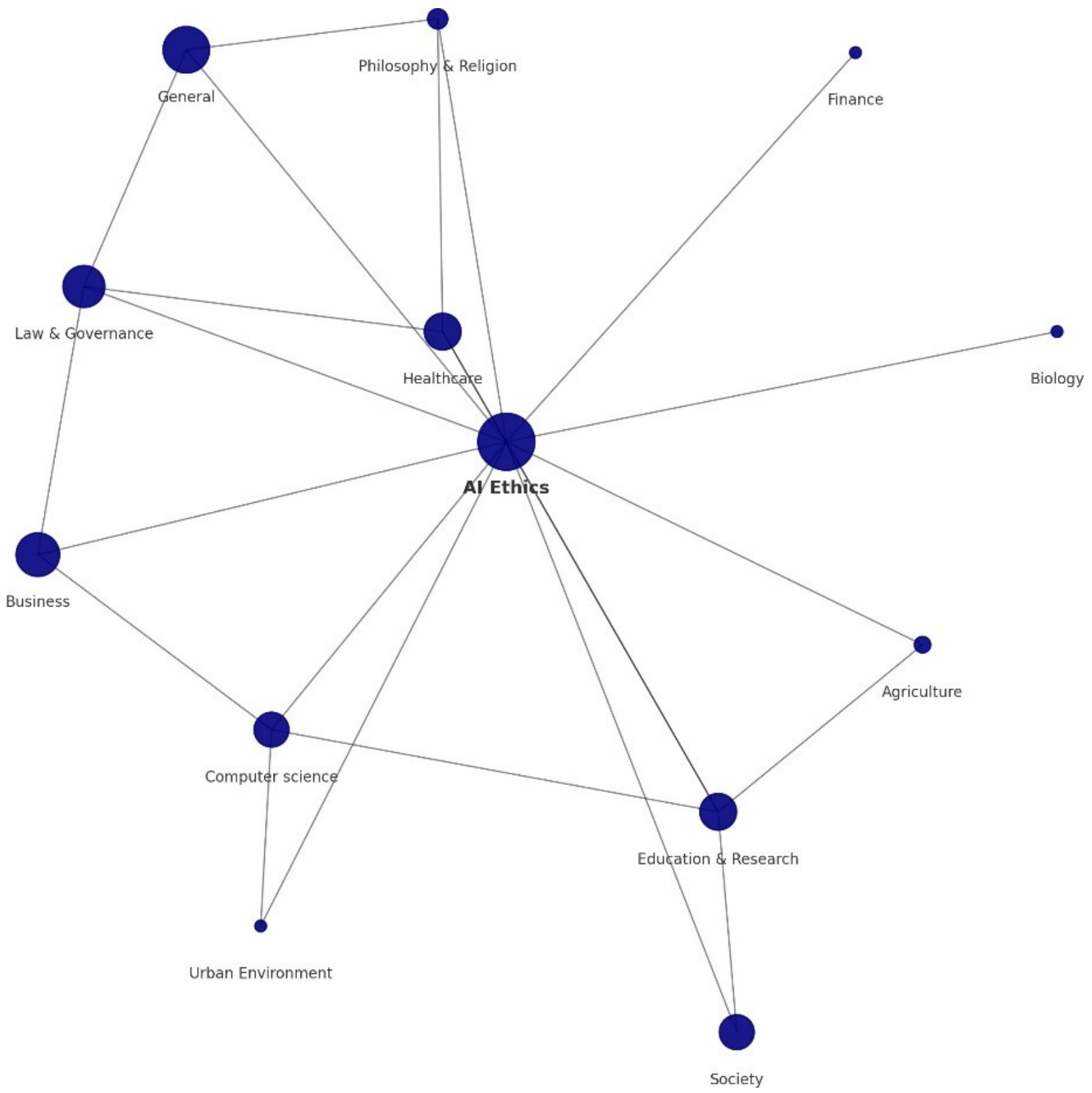

Fig. 10. Classification of publications by domains

The *General* domain, which constitutes the most extensive category within our dataset, encompasses a diverse array of themes, with its primary objective being to offer a comprehensive overview of the AI Ethics landscape [17]. Key topics within this domain include the exploration of ethical principles [31] and their associated trade-offs [44], such as *transparency* versus *privacy*, and the formulation of guidelines to navigate these complex ethical terrains [19]. Additionally, discussions within this domain extend to the broader examination of risks and challenges associated with AI technologies [45].

Among the specific fields analyzed, *Business* domain emerged as the most extensively studied. The primary focus in this domain was on bridging the gap between theoretical concepts and practical applications [27]. Key discussions revolve around how ethical principles can be operationalized in real-world scenarios. This includes exploring the implications of AI in the workplace, particularly in relation to employee safety [15, 16]. A significant topic in this domain is the use of AI in hiring processes [51], examining both the potential benefits and the ethical concerns, such as *bias* and *fairness*. Studies in this domain essentially shed light on the challenges and opportunities of integrating AI ethically in a corporate environment, which seems to be the main concern in the current scientific literature.

In the realm of *Education & Research*, our analysis indicates a growing emphasis on integrating AI Ethics into academic curricula. The studies in this domain delve into the rationale and methodologies for incorporating AI Ethics into high school [34] and university programs [24]. Another salient theme within this domain is the use of AI for surveillance and online proctoring in educational settings [18, 21]. These studies critically assess the advantages and drawbacks of AI-enabled applications, exploring both the effectiveness of these technologies in maintaining academic integrity and their impact on student privacy and perceptions. *Education & Research* domain, therefore, encompasses not only the content of AI Ethics education but also the ethical dimensions of AI applications in the educational process and research practice itself.

*Healthcare* domain focuses on the ethical dimensions surrounding the development and use of AI in a medical context. A primary theme here involves evaluating the risks and challenges associated with AI-enabled medical devices [23] and products [42]. Data protection and patient privacy emerge as prevalent topics [13], reflecting the heightened sensitivity of personal health information in the age of AI. Additionally, the use and implications of conversational AI, such as chatbots, in healthcare are explored [8]. Collectively, the *Healthcare* domain addresses the multifaceted ethical issues at the intersection of AI technology and patient care, with an emphasis on personal data protection.

*Law & Governance* domain, often encountered as a secondary topic, is mostly centered around data privacy under the GDPR framework [11, 39]. A substantial portion of the discussion also focuses on the existing regulatory initiatives of different countries and the necessity for robust governance to address the ethical challenges posed by AI[19]. The studies in this domain critically examine the gaps in the regulatory frameworks and advocate for more comprehensive policies. *Law & Governance* domain underscores the importance of legal structures in safeguarding ethical standards in AI.

In *Society* domain the focal point is predominantly on public opinion towards AI. Studies in this domain explore societal perspectives on AI and how the public perceives the importance of various ethical principles [33]. The main focus is on the general attitudes towards AI [28], including concerns and expectations, and public reactions to the potential misuse of AI [47]. By capturing the societal sentiment towards AI, the studies in this domain shed light on the ethical dimensions of AI as viewed through the lens of the broader public. This perspective is vital for understanding the societal impact of AI and for guiding responsible AI development that aligns with public values and concerns.

*Computer Science* domain, the last significantly covered field in the reviewed literature, delves into the ethical considerations specific to various AI technologies. Key themes include the ethics of particular technologies like emotion recognition systems [40], where the moral implications of technology use and its impact on privacy and autonomy are examined. Additionally, the domain encompasses the study of ethical design principles, especially in the development of AI-driven tools such as chatbots [12], focusing on how these technologies can be designed to uphold ethical standards. Cybersecurity also emerges as a critical topic, particularly in the context of adversarial attacks [32], where the integrity and resilience of AI systems against malicious interventions are scrutinized. This domain, therefore, intersects the technical aspects of AI with ethical deliberation, emphasizing the need for ethical mindfulness in the technological design and deployment of AI.

Domains such as *Agriculture*, *Biology*, *Finance*, and *Urban Environment* are underrepresented in our dataset, possibly due to their specific niche applications or the nascent stage of AI Ethics in these areas. These fields might be overshadowed by sectors where AI's impact is more immediate and visible to society, leading to a disproportionate focus in current research. In addition, *Philosophy & Religion* domain, while being fundamental in framing broader ethical questions, seems to be less significant than more practical concerns in other domains.

*4.4.3 AI Ethics Frameworks.* Among the 73 studies analyzed, a limited fraction—slightly over 13 percent—placed primary emphasis on the development of framework for AI Ethics.

Frameworks discovered within *General* domain exhibit a predominantly theoretical focus. For instance, Ashok et al. [10] developed an ontological framework aimed at conceptualizing the ethical impacts of AI across physical, cognitive, information, and governance domains. This framework provides a high-level theoretical overview rather than actionable steps for ethical AI implementation. Similarly, Sharma et al. [48] proposed a framework for evaluating ethics in AI centered around 7 abstract checkpoints, which, while insightful, does not translate easily into practical measures.

Within *Healthcare* domain, a single framework was identified, which explores the risks associated with AI-enabled chatbots in the context of medical ethics [20]. The study's framework, although adeptly maps ethical principles to potential risks, stops short of offering a structured risk management approach.

A more applied approach is evident in *Business* domain, with several frameworks designed to integrate ethical considerations into the corporate landscape. These include methodologies for assessing workplace health and safety in the context of AI [15, 16]. Agbese et al. propose a novel way of implementing ethical requirements in organizations by utilizing the Agile portfolio management framework [6]. It is worth noting, however, that this framework primarily concentrates on the strategic and planning aspects of companies, emphasizing the integration of ethical requirements rather than specific ethical aspects that should be considered. Meanwhile, another notable study in this domain introduces an AI maturity model for small and medium sized companies (also known as SME) [46]. While this model proves to be practical and has undergone validation by real organizations, its main function lies in providing an initial assessment of companies, offering limited insights and strategies for improvement.

ECCOLA method by Vakkuri et al. represents a beacon of practical application within our dataset, employing a deck of cards to facilitate ethical considerations in AI systems [50]. It is "a sprint-by-sprint" process designed to facilitate ethical thinking in AI and autonomous systems development that demonstrates a tangible approach to embedding ethical thinking within organizational processes. ECCOLA comprises a set of 21 cards, split into 8 distinct themes. These themes include *analyze, transparency, safety & security, fairness, data, agency & oversight, wellbeing, and account- ability*, with each theme encompassing 1 to 6 cards. The employed methodology is covering all key ethical principles and providing organizations with a practical tool for implementing ethically aligned AI systems.

Upon reviewing AI Ethics frameworks across different fields, including the *General* domain and specific sectors like *Healthcare* and *Business*, a discernible pattern emerges beyond the unique context of each study. Despite diverse applications and focus areas, four consistent dimensions come to the forefront: ***Data, Technology, Process***, and ***People***. These dimensions reoccur in frameworks regardless of their specific field of application, suggesting a potentially universal approach within the realm of AI Ethics.

To sum up, our review, aimed at uncovering frameworks with a strong focus on AI privacy and security, identified only one study (ECCOLA methodology) that partially addressed our inquiry. The scarcity of frameworks that effectively meld ethical theory with actionable practice highlights a noticeable gap in the existing scientific literature. Despite the heightened public and scholarly concern regarding privacy and security in AI systems, among the studies scrutinized, none provided a framework dedicated exclusively to the practical evaluation of privacy or security aspects in AI models.

*4.4.4 Privacy and Security in AI Ethics.* The realms of privacy and security within the field of AI Ethics have garnered substantial attention in the scientific literature. Our examination reveals a pronounced focus on privacy considerations (64.38%), often anchored in compliance with the General Data Protection Regulation (GDPR).

The majority of studies underscore the profound influence of GDPR in shaping discussions on privacy within the AI landscape[9, 11]. As a comprehensive data protection regulation, GDPR stands as a cornerstone guiding how AI systems should handle data collection, processing, and usage. These studies delve into the nuanced implications of GDPR compliance, highlighting the imperative for AI developers and practitioners to align their systems with the stringent privacy requirements outlined in the regulation.

Another recurring theme across these studies is the imperative for AI technologies to align with fundamental human rights[36, 51]. As AI becomes increasingly integrated into diverse facets of society, the ethical discourse extends beyond technical considerations to encompass the profound impact on human lives. The studies consistently emphasize the critical need for AI development and deployment to uphold and respect universally recognized human rights principles. This encompasses the right to privacy, freedom of expression, non-discrimination, and the right to fair and unbiased treatment.

Our analysis reveals that both privacy and security considerations coexist in 32.88% of the selected studies. This indicates a tendency to discuss these domains together or within the broader context of AI Ethics principles. Security considerations alone are present in only 2.74%, showcasing a clear lack of attention towards security topics in the current literature.

The above-mentioned trends in the selected studies emphasize the importance of AI governance. The evolving landscape of AI technologies necessitates a governance framework ensuring responsible and ethical AI development and deployment. AI governance emerges as a key enabler in navigating the intricate challenges posed by privacy and security considerations.

A notable development in this regard is the European Union's proposal of the EU AI Act,

representing a groundbreaking initiative as the first regulation specifically focused on artificial intelligence. Adopting a risk-based approach, the EU AI Act classifies AI systems into distinct risk categories, delineating corresponding requirements and obligations. This legislative effort underscores a commitment to addressing ethical and societal impacts of AI at the legal level, serving as a pioneering step in shaping the future landscape of AI law.

In summary, the selected studies underscore the symbiotic relationship between privacy and security considerations in the AI domain, with the evident prevalence of privacy discussions in the current scientific discourse. The analysis indicates a high correlation between the existing AI governance initiatives and the degree of concern towards the ethical implementation of AI.

## 5 TOWARDS PRIVACY- AND SECURITY-AWARE FRAMEWORK FOR ETHICAL AI

Developing a privacy- and security-aware framework for Ethical AI requires a multifaceted approach. The systematic literature review has identified four pivotal dimensions - Data, Technology, People, and Process. These dimensions will serve as the foundational pillars of our framework, guiding the ethical assessment of AI systems in organizational environments.

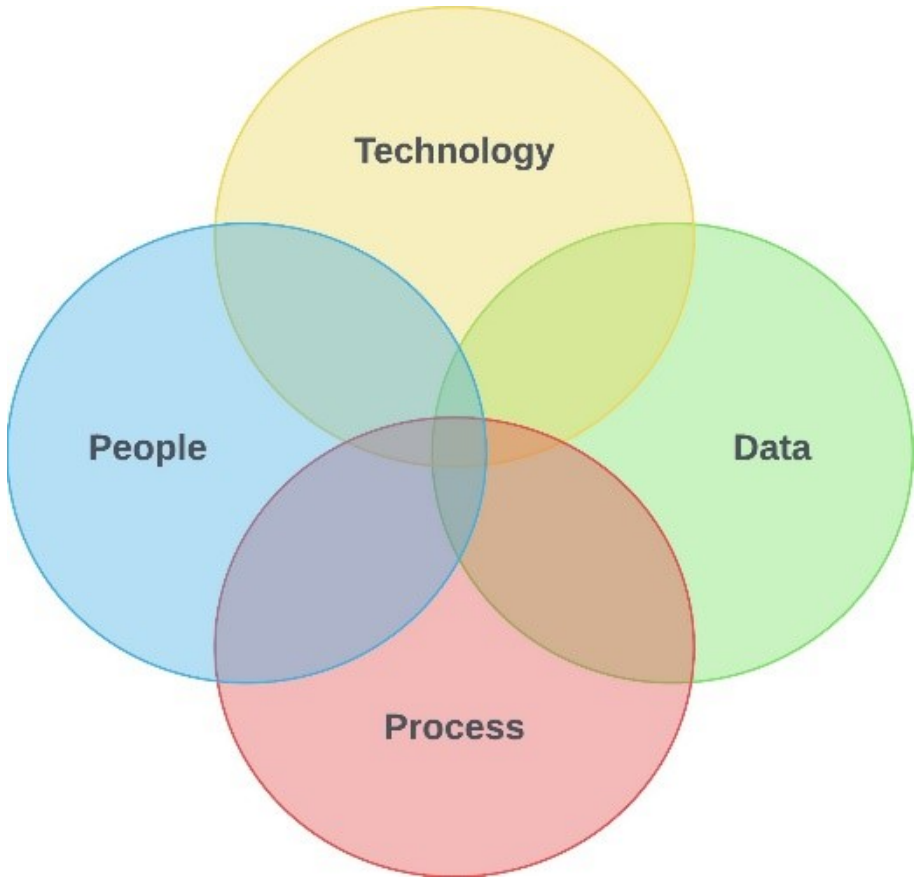

**Fig. 11.** Four pillars of AI systems

**Data dimension** is foundational, recognizing that data is the substrate from which AI derives its functionality and intelligence, i.e., by which it is fueled. Ethical considerations in data governance, such as consent, anonymization, and data rights, are paramount, particularly as they relate to privacy and the potential for misuse.

***Recommendation 1:*** *Organizations should mandate rigorous scrutiny of data handling practices to mitigate risks and protect individual rights.*

**Technology dimension** addresses the tangible components of AI systems. The security mechanisms and technical safeguards are not merely adjuncts but are intrinsic to the ethical fabric of the technology. This necessitates an ongoing commitment to adapt and refine these measures, in line with the latest technological advancements.

***Recommendation 2:*** *Organizations should adopt a 'security-by-design' approach and continuously align with security standards, ensuring that security mechanisms are foundational elements in the development of AI.*

The human element, encapsulated in **People dimension**, underscores societal and individual factors that influence AI systems. This dimension prioritizes secure authentication and permission management. Additionally, it advocates for active engagement with a diverse array of stakeholders, ensuring that AI practices are informed by a broad spectrum of ethical perspectives and experiences.

***Recommendation 3:*** *Organizations should establish robust protocols for authentication and permission management.*
***Recommendation 4:*** *Organizations should foster awareness of AI privacy and security concerns and nurture the practical skills necessary to address these challenges effectively.*

**Process dimension** is crucial in materializing ethical principles into tangible actions. It **involves** instituting robust processes for ethical oversight, from compliance checks to ethical auditing and incident response strategies, ensuring that AI systems adhere to established standards throughout their lifecycle. This dimension captures the operationalization of ethical principles through established procedures and includes scrutinizing the extent to which these processes are ingrained in the organization's culture and AI system management.

***Recommendation 5:*** *Organizations should establish and rigorously enforce comprehensive governance processes, encompassing ethical oversight, compliance checks, and incident response strategies.*

For our privacy- and security-aware framework we synthesized 4 dimensions derived from the SLR – Data, Technology, People, and Process – into an integrated structure, as illustrated in Table 6, enabling a grounded approach to AI ethics.

Each dimension is guided by a set of specific questions to encompass the overarching themes of privacy and security within AI systems. Data, People, and Process dimensions are each guided by three questions, while Technology dimension is informed by two questions. These questions are deliberately wide-ranging, designed to provoke thorough contemplation and exploration of ethical practices. They were formulated based on the SLR-established domain knowledge, ensuring that they are deeply rooted in the current understanding and challenges of AI ethics.

Table 6. Privacy- and Security-Aware Framework for AI Ethics

| Dimension | Question | Best Practice | Risk Level |
|---|---|---|---|
| **Data** | How do you classify your data? | Identify the data | High |
| | | Define data protection controls | High |
| | | Define data lifecycle management | Low |
| | How do you manage data privacy? | Implement consent management | High |
| | | Utilize anonymization techniques | High |
| | | Define data rights and ownership policies | Medium |
| | How do you protect your data? | Implement secure key management | High |
| | | Implement encryption | High |
| | | Implement access control | Medium |
| | | Utilize tools to prevent direct access to data | Low |
| **Technology** | How do you protect your resources? | Perform vulnerability management | High |
| | | Control network traffic at all | High |
| | | layers | |
| | | Reduce attack surface | High |
| | | Validate software integrity | Medium |
| | How do you secure your applications? | Perform regular penetration testing | High |
| | | Assess security of pipelines | High |
| | | Deploy software programmatically | High |
| | | Centralize services for packages and dependencies | Medium |
| **People** | How do you manage authentication? | Utilize strong sign-in mechanisms | High |
| | | Utilize temporary credentials | High |
| | | Audit and rotate credentials periodically | High |
| | | Store and use secrets securely | High |
| | How do you manage access control? | Determine access requirements | High |
| | | Follow the principle of least privilege | High |
| | | Manage access based on lifecycle | Low |
| | | Share resources securely | Medium |
| | How do you ensure stakeholder engagement? | Facilitate regular interactive privacy and security trainings | High |
| | | Implement stakeholder feedback mechanisms | Medium |
| **Process** | How do you ensure regulatory compliance? | Implement regular compliance audits | High |
| | | Perform regulatory impact assessments | High |
| | How do you detect potential incidents? | Configure logging | High |
| | | Analyze metrics and logs centrally | High |
| | | Automate monitoring of critical events | Medium |
| | How do you manage incidents? | Identify key personnel and resources | High |
| | | Create incident management plans | High |
| | | Facilitate learning from incidents | Medium |

To enhance the framework's utility, we elaborated a suite of best practices corresponding to each question, offering organizations targeted guidance for ethical assessment. Data and People dimensions incorporate 10 best practices each, while Technology and Process dimensions are outlined with 8 best practices. This methodical compilation of best practices serves as a concrete reference point for organizations seeking to evaluate and improve their AI systems.

Embracing the complexity of ethical considerations, we adopted a risk-based approach utilized in EU AI Act. This is an approach proposed by the European Commission to regulate AI systems by categorizing them into four levels: *Unacceptable Risk*, *High Risk*, *Limited Risk*, and *Minimal or No Risk*. AI systems associated with unacceptable risk pose direct threats to public safety and fundamental rights, meriting their outright prohibition. This is followed by high-risk AI systems, employed in areas such as critical infrastructure and healthcare, where the potential for significant impacts on safety and rights necessitates rigorous regulatory adherence. Limited-risk AI systems, like chatbots, require explicit transparency measures, ensuring user awareness of AI interaction. The spectrum concludes with minimal or no risk AI systems, like most AI-enabled video games or spam filters, which are subject to minimal regulatory oversight [2].

In our framework, we've mirrored this risk-based approach, focusing on *High*, *Limited*, and *Minimal* risk categories, aligning our best practices in a manner that reflects the level of potential impact, defined as *High*, *Medium*, or *Low*. This provides us with the possibility to critically evaluate and manage ethical risks of AI. Our framework strategically excluded unacceptable risks, as systems falling under this classification are not only prohibited by the AI Act but also represent a class of AI development that fundamentally contradicts ethical principles and standards.

The proposed framework indicates the prevalence of practices potentially leading to high-risk impact across all dimensions. This includes practices such as implementing consent management in Data dimension and performing vulnerability management within Technology dimension. People dimension contains the highest count of practices with high-risk impact (n = 7), while the other three dimensions have equal numbers (n = 6). This underscores the importance of human-centric considerations in AI. Practices with medium-risk impact are equally distributed across dimensions (n = 2), and low-risk impact is present only in Data and People dimensions.

This comprehensive analysis of the dimensions, practices, and associated risk levels highlights the framework's capacity to navigate the complex terrain of AI ethics. It showcases the emphasis on a balanced approach, where both the micro-level intricacies and macro-level perspectives are equally considered, ensuring that the deployment of AI systems is both pragmatic and grounded in a robust understanding of the evolving landscape.

## 6 DISCUSSION AND LIMITATIONS

This study has introduced a Privacy- and Security-Aware Framework for Ethical AI, developed based on the conceptualized findings of the SLR. The framework, centered around the critical dimensions of *Data, Technology, People*, and *Process*, is designed to guide organizations in evaluating and enhancing their AI systems' privacy and security postures. By addressing key questions within each dimension, we propose a series of best practices accompanied by risk assessments to inform the decision-making.

The integration of findings from our SLR has been instrumental in shaping this framework. The SLR underscored the escalating concerns regarding AI's privacy and security aspects and revealed a gap in practical, actionable frameworks addressing these issues. Aligning our framework with these findings equips it with the acumen to effectively address these concerns, offering organizations a clear path to enhance AI ethics compliance and practice.

While the framework aims to be comprehensive, certain limitations must be acknowledged. The SLR that provided the underlying data for this framework has inherent limitations. The exclusive selection of studies from 2020 to 2023 and reliance on a single database may narrow the framework's scope, potentially omitting seminal works and diverse perspectives. The targeted search strategy, focusing on the intersection of privacy, security, and AI Ethics, ensures the depth of the research but could also miss broader ethical concerns.

To date, the framework has not been empirically tested within the organizational environment. As a result, a key aspect of future work involves validation of the proposed theoretical and therefore conceptual framework, which will provide empirical data to refine it further and adapt to diverse organizational contexts. It is also imperative to consider the scalability of the framework, ensuring that it can be effectively utilized by organizations of varying sizes and capacities.

Moreover, the framework assumes a certain level of organizational maturity in privacy and security practices, which may not be present in all organizations. There can also be industry-specific considerations that the framework does not fully encapsulate, necessitating customization in certain sectors. The framework might also benefit from incorporating stakeholders into its structure, as it would further operationalize it.

The rapid pace of technological innovation and the continuous evolution of legal standards mean that the framework will need ongoing updates to remain relevant and maintain compliance.

Acknowledging these constraints is not to undermine the framework's value but to pave the way for its continuous improvement. As AI Ethics is an ever-evolving field, the framework must adapt and expand, incorporating a wider array of sources and perspectives to remain both current and comprehensive.

## 7 CONCLUSION

The development and introduction of the Privacy- and Security-Aware Framework for Ethical AI signify a critical advance in the field of AI Ethics. This framework, with its emphasis on the dimensions of Data, Technology, People, and Process, offers a systematic method for organizations to assess and enhance their AI systems, ensuring that privacy and security are not ancillary considerations but central to the ethical deployment of AI technologies.

Our work initiated with a systematic literature review, shedding light on prevalent concerns in both public and academic discussions regarding privacy and security in AI. This review laid the foundation for a framework that bridges the gap between theoretical ethical concepts and their practical application. By integrating elements such as risk assessments and best practices tailored to each dimension, we have developed a tool that responds to the urgent need for structured guidance in navigating the ethical complexities of AI.

The practical implications of this framework are far-reaching. It can serve as a blueprint for organizations to evaluate and fortify the ethical integrity of their AI systems, focusing particularly on privacy and security aspects. We encourage organizations to utilize this framework as a guide for ethical decision-making, thereby fostering an environment where AI is developed and utilized with a strong commitment to ethical principles.

The subsequent phase of this research will involve implementing the framework in a variety of organizational settings. This practical application is expected to yield valuable insights, allowing for the refinement of the framework and its adaptation to a wide range of industries and organizational sizes. It is anticipated that this will also uncover new challenges and ethical considerations,

Looking ahead, the framework is positioned for ongoing evolution. We recognize the necessity of adapting to emerging technologies and evolving ethical standards. As the AI landscape continually transforms, so must our approaches to ethical governance. The Privacy- and Security-Aware Framework for AI Ethics marks the beginning of a sustained endeavor in ethical discourse, refinement, and practical implementation. The ultimate goal of our research is to create a paradigm where ethical considerations are seamlessly integrated into the fabric of AI development and deployment.

## REFERENCES

[1] [n. d.]. *Ethics guidelines for trustworthy AI | Shaping Europe's digital future*. https://digital-strategy.ec.europa.eu/en/library/ethics-guidelines-trustworthy-ai

[2] [n. d.]. *EU AI Act: first regulation on artificial intelligence | News | European Parliament.* https://www.europarl.europa. eu/news/en/headlines/society/20230601STO93804/eu-ai-act-first-regulation-on-artificial-intelligence

[3] 2023. AI and Privacy: The privacy concerns surrounding AI, its potential impact on personal data. *The Economic Times* (April 2023). https://economictimes.indiatimes.com/news/how-to/ai-and-privacy-the-privacy-concerns-surrounding-ai-its-potential-impact-on-personal-data/articleshow/99738234.cms

[4] 2023. *Artificial Intelligence Act: deal on comprehensive rules for trustworthy AI | News | European Parlia- ment.* https://www.europarl.europa.eu/news/en/press-room/20231206IPR15699/artificial-intelligence-act-deal-on- comprehensive-rules-for-trustworthy-ai

[5] C. Adams, P. Pente, G. Lemermeyer, and G. Rockwell. 2023. Ethical principles for artificial intelligence in K-12 education. *Computers and Education: Artificial Intelligence* 4 (2023). https://doi.org/10.1016/j.caeai.2023.100131

[6] M. Agbese, R. Mohanani, A. Khan, and P. Abrahamsson. 2023. Implementing AI Ethics: Making Sense of the Ethical Requirements. In *ACM Int. Conf. Proc. Ser.* Association for Computing Machinery, 62–71. https://doi.org/10.1145/3593434.3593453

[7] K. Ahmad, M. Maabreh, M. Ghaly, K. Khan, J. Qadir, and A. Al-Fuqaha. 2022. Developing future human-centered smart cities: Critical analysis of smart city security, Data management, and Ethical challenges. *Computer Science Review* 43 (2022). https://doi.org/10.1016/j.cosrev.2021.100452

[8] R. Ahmad, D. Siemon, U. Gnewuch, and S. Robra-Bissantz. 2021. The benefits and caveats of personality-adaptive conversational agents in mental health care. In *Annu. Americas Conf. Inf. Syst., AMCIS*. Association for Information Systems. https://www.researchgate.net/publication/351659439_The_Benefits_and_Caveats_of_Personality-Adaptive_Conversational_Agents_in_Mental_Health_Care

[9] Mohammad Mohammad Amini, Marcia Jesus, Davood Fanaei Sheikholeslami, Paulo Alves, Aliakbar Hassanzadeh Benam, and Fatemeh Hariri. 2023. Artificial Intelligence Ethics and Challenges in Healthcare Applications: A Comprehensive Review in the Context of the European GDPR Mandate. *Mach. Learn. Knowl. Extr.* 5, 3 (Sept. 2023), 1023–1035. https://doi.org/make5030053

[10] M. Ashok, R. Madan, A. Joha, and U. Sivarajah. 2022. Ethical framework for Artificial Intelligence and Digital technologies. *International Journal of Information Management* 62 (2022). https://doi.org/10.1016/j.ijinfomgt.2021.102433

[11] D. Banciu and C.E. Cirnu. 2022. AI Ethics and Data Privacy compliance. In *Int. Conf. Electron., Comput. Artif. Intell., ECAI.* Institute of Electrical and Electronics Engineers Inc. https://doi.org/10.1109/ECAI54874.2022.9847510

[12] J. Bang, S. Kim, J.W. Nam, and D.-G. Yang. 2021. Ethical Chatbot Design for Reducing Negative Effects of Biased Data and Unethical Conversations. In *Int. Conf. Platf. Technol. Serv., PlatCon - Proc.* Institute of Electrical and Electronics Engineers Inc. https://doi.org/10.1109/PlatCon53246.2021.9680760

[13] J. Baric-Parker and E.E. Anderson. 2020. Patient Data-Sharing for AI: Ethical Challenges, Catholic Solutions. *Linacre Quarterly* 87, 4 (2020), 471–481. https://doi.org/10.1177/0024363920922690

[14] C. Cadwalladr and E. Graham-Harrison. 2018. Revealed: 50 million Facebook profiles harvested for Cambridge Analytica in major data breach. *The Guardian* (March 2018). https://www.theguardian.com/news/2018/mar/17/cambridge-analytica-facebook-influence-us-election

[15] A. Cebulla, Z. Szpak, C. Howell, G. Knight, and S. Hussain. 2023. Applying ethics to AI in the workplace: the design of a scorecard for Australian workplace health and safety. *AI and Society* 38, 2 (2023), 919–935. https://doi.org/10.1007/s00146-022-01460-9

[16] A. Cebulla, Z. Szpak, and G. Knight. 2023. Preparing to work with artificial intelligence: assessing WHS when using AI in the workplace. *International Journal of Workplace Health Management* 16, 4 (2023), 294–312. https://doi.org/10.1108/IJWHM-09-2022-0141

[17] M. Christoforaki and O. Beyan. 2022. AI Ethics—A Bird's Eye View. *Applied Sciences (Switzerland)* 12, 9 (2022). https://doi.org/10.3390/app12094130

[18] S. Coghlan, T. Miller, and J. Paterson. 2021. Good Proctor or "Big Brother"? Ethics of Online Exam Supervision Technologies. *Philosophy and Technology* 34, 4 (2021), 1581–1606. https://doi.org/10.1007/s13347-021-00476-1

[19] N. Corrêa, C. Galvão, J.W. Santos, C. Del Pino, E.P. Pinto, C. Barbosa, D. Massmann, R. Mambrini, L. Galvão, E. Terem, and N. de Oliveira. 2023. Worldwide AI ethics: A review of 200 guidelines and recommendations for AI governance. *Patterns* 4, 10 (2023). https://doi.org/10.1016/j.patter.2023.100857

[20] E. Fournier-Tombs and J. McHardy. 2023. A Medical Ethics Framework for Conversational Artificial Intelligence. *Journal of Medical Internet Research* 25 (2023). https://doi.org/10.2196/43068

[21] B. Han, G. Buchanan, and D. McKay. 2022. Learning in the Panopticon: Examining the Potential Impacts of AI Monitoring on Students. In *ACM Int. Conf. Proc. Ser.* Association for Computing Machinery, 9–21.

https://doi.org/10. 1145/3572921.3572937

[22] Jeonghye Han. [n. d.]. An Information Ethics Framework Based on ICT Platforms. 13, 9 ([n. d.]), 440. https://doi.org/10.3390/info13090440

[23] Jonathan Herington, Melissa D. McCradden, Kathleen Creel, Ronald Boellaard, Elizabeth C. Jones, Abhinav K. Jha, Arman Rahmim, Peter J. H. Scott, John J. Sunderland, Richard L. Wahl, Sven Zuehlsdorff, and Babak Saboury. [n. d.]. Ethical Considerations for Artificial Intelligence in Medical Imaging: Data Collection, Development, and Evaluation. 64, 12 ([n. d.]), 1848–1854. https://doi.org/10.2967/jnumed.123.266080

[24] A. Hingle and A. Johri. 2023. Recognizing Principles of AI Ethics through a Role-Play Case Study on Agriculture. In *ASEE Annu. Conf. Expos. Conf. Proc.* American Society for Engineering Education. https://peer.asee.org/44029

[25] H. Hirsch-Kreinsen and T. Krokowski. 2023. Trustworthy AI: AI made in Germany and Europe? *AI and Society* (2023). https://doi.org/10.1007/s00146-023-01808-9

[26] C. Huang, Z. Zhang, B. Mao, and X. Yao. 2023. An Overview of Artificial Intelligence Ethics. *IEEE Transactions on Artificial Intelligence* 4, 4 (2023), 799–819. https://doi.org/10.1109/TAI.2022.3194503

[27] J. Ibáñez and M. Olmeda. 2022. Operationalising AI ethics: how are companies bridging the gap between practice and principles? An exploratory study. *AI and Society* 37, 4 (2022), 1663–1687. https://doi.org/10.1007/s00146-021-01267-0

[28] Y. Ikkatai, T. Hartwig, N. Takanashi, and H.M. Yokoyama. 2022. Octagon Measurement: Public Attitudes toward AI Ethics. *International Journal of Human-Computer Interaction* 38, 17 (2022), 1589–1606. https://doi.org/10.1080/10447318.2021.2009669

[29] Anna Jobin, Marcello Ienca, and Effy Vayena. 2019. The global landscape of AI ethics guidelines. *Nature Machine Intelligence* 1 (Sept. 2019). https://doi.org/10.1038/s42256-019-0088-2

[30] M.K. Kamila and S.S. Jasrotia. 2023. Ethical issues in the development of artificial intelligence: recognizing the risks. *International Journal of Ethics and Systems* (2023). https://doi.org/10.1108/IJOES-05-2023-0107 Publisher: Emerald Publishing.

[31] A. Khan, M.A. Akbar, M. Fahmideh, P. Liang, M. Waseem, A. Ahmad, M. Niazi, and P. Abrahamsson. 2023. AI Ethics: An Empirical Study on the Views of Practitioners and Lawmakers. *IEEE Transactions on Computational Social Systems* (2023), 1–14. https://doi.org/10.1109/TCSS.2023.3251729

[32] M. Khosravy, K. Nakamura, A. Pasquali, O. Witkowski, N. Nitta, and N. Babaguchi. 2022. When AI Facilitates Trust Violation: An Ethical Report on Deep Model Inversion Privacy Attack. In *Proc. - Int. Conf. Comput. Sci. Comput. Intell., CSCI*. Institute of Electrical and Electronics Engineers Inc., 929–935. https://doi.org/10.1109/CSCI58124.2022.00166

[33] K. Kieslich, B. Keller, and C. Starke. 2022. Artificial intelligence ethics by design. Evaluating public perception on the importance of ethical design principles of artificial intelligence. *Big Data and Society* 9, 1 (2022). https://doi.org/10.1177/20539517221092956

[34] Z. Kilhoffer, Z. Zhou, F. Wang, F. Tamton, Y. Huang, P. Kim, T. Yeh, and Y. Wang. 2023. 'How technical do you get? I'm an English teacher': Teaching and Learning Cybersecurity and AI Ethics in High School. In *Proc. IEEE Symp. Secur. Privacy*, Vol. 2023-May. Institute of Electrical and Electronics Engineers Inc., 2032–2049. https://doi.org/10.1109/SP46215.2023.10179333

[35] B. Kitchenham. 2004. Procedures for Performing Systematic Reviews. *Keele, UK, Keele Univ.* 33 (08 2004). https://www.researchgate.net/publication/228756057_Procedures_for_Performing_Systematic_Reviews

[36] A. Kriebitz and C. Lütge. 2020. Artificial Intelligence and Human Rights: A Business Ethical Assessment. *Business and Human Rights Journal* 5, 1 (2020), 84–104. https://doi.org/bhj.2019.28

[37] Sourabh Kulesh and Nikunj Dalmia. [n. d.]. In the public AI: Big tech giants like Microsoft, Meta, Google experiment with artificial intelligence tools. ([n. d.]). https://economictimes.indiatimes.com/magazines/panache/in-the-public-ai-big-tech-giants-like-microsoft-meta-google-experiment-with-artificial-intelligence-tools/articleshow/102869083.cms?from=mdr

[38] B.M. McLaren. [n. d.]. Extensionally defining principles and cases in ethics: An AI model. 150, 1 ([n. d.]), 145–181. https://doi.org/10.1016/S0004-3702(03)00135-8

[39] M. Milossi, E. Alexandropoulou-Egyptiadou, and K.E. Psannis. 2021. AI Ethics: Algorithmic Determinism or Self-Determination? The GPDR Approach. *IEEE Access* 9 (2021), 58455–58466. https://doi.org/10.1109/ACCESS.2021.3072782

[40] S.M. Mohammad. 2022. Ethics Sheet for Automatic Emotion Recognition and Sentiment Analysis. *Computational Linguistics* 48, 2 (2022), 239–278. https://doi.org/10.1162/coli_a_00433

[41] M. Nishanthi, H. U. C. S. Kumara, and K. W. a. M. Konpola. [n. d.]. Research Conception of Palm Leaf Manuscript Conservation: Bibliometric Analysis of Scopus database. 10, 2 ([n. d.]), 13–28. https://journals.sjp.ac.lk/index.php/ijms/article/view/6477

[42] S. Pasricha. 2022. AI Ethics in Smart Healthcare. *IEEE Consumer Electronics Magazine* (2022), 1–7.

https://doi.org/10. 1109/MCE.2022.3220001

[43] E. Pilkington. 2019. Google's secret cache of medical data includes names and full details of millions – whistleblower. *The Guardian* (Nov. 2019). https://www.theguardian.com/technology/2019/nov/12/google-medical-data-project-nightingale-secret-transfer-us-health-information

[44] C. Sanderson, D. Douglas, and Q. Lu. 2023. Implementing Responsible AI: Tensions and Trade-Offs Between Ethics Aspects. In *Proc Int Jt Conf Neural Networks*, Vol. 2023-June. Institute of Electrical and Electronics Engineers Inc. https://doi.org/10.1109/IJCNN54540.2023.10191274

[45] C. Sanderson, Q. Lu, D. Douglas, X. Xu, L. Zhu, and J. Whittle. 2022. Towards Implementing Responsible AI. In *Proc. - IEEE Int. Conf. Big Data, Big Data*. Institute of Electrical and Electronics Engineers Inc., 5076–5081. https: //doi.org/10.1109/BigData55660.2022.10021121

[46] T. Schuster and L. Waidelich. 2022. Maturity of Artificial Intelligence in SMEs: Privacy and Ethics Dimensions. In *IFIP Advances in Information and Communication Technology*, Vol. 662. Springer Science and Business Media Deutschland GmbH, 274–286. https://doi.org/10.1007/978-3-031-14844-6_22

[47] D.B. Shank and A. Gott. 2020. Exposed by AIs! People Personally Witness Artificial Intelligence Exposing Personal Information and Exposing People to Undesirable Content. *International Journal of Human-Computer Interaction* 36, 17 (2020), 1636–1645. https://doi.org/10.1080/10447318.2020.1768674

[48] V. Sharma, N. Mishra, V. Kukreja, A. Alkhayyat, and A.A. Elngar. 2023. Framework for Evaluating Ethics in AI. In *Int. Conf. Innov. Data Commun. Technol. Appl., ICIDCA - Proc.* Institute of Electrical and Electronics Engineers Inc., 307–312. https://doi.org/10.1109/ICIDCA56705.2023.10099747

[49] K. Siau and W. Wang. 2020. Artificial intelligence (AI) Ethics: Ethics of AI and ethical AI. *Journal of Database Management* 31, 2 (2020), 74–87. https://doi.org/10.4018/JDM.2020040105

[50] V. Vakkuri, K.-K. Kemell, M. Jantunen, E. Halme, and P. Abrahamsson. 2021. ECCOLA — A method for implementing ethically aligned AI systems. *Journal of Systems and Software* 182 (2021). https://doi.org/10.1016/j.jss.2021.111067

[51] J. Yam and J.A. Skorburg. 2021. From human resources to human rights: Impact assessments for hiring algorithms. *Ethics and Information Technology* 23, 4 (2021), 611–623. https://doi.org/10.1007/s10676-021-09599-7

[52] Y. Zhang, M. Wu, G.Y. Tian, G. Zhang, and J. Lu. 2021. Ethics and privacy of artificial intelligence: Understandings from bibliometrics. *Knowledge-Based Systems* 222 (2021). https://doi.org/10.1016/j.knosys.2021.106994

[53] A. Zuiderwijk, Y.C. Chen, and F. Salem. [n. d.]. Implications of the use of artificial intelligence in public governance: A systematic literature review and a research agenda. 38, 3 ([n. d.]), 101577. https://doi.org/10.1016/j.giq.2021.101577